\documentclass{rmf-d}
\usepackage{nopageno,rmfbib,multicol,times,epsf,amsmath,amssymb,cite}
\usepackage[T1]{fontenc} 
\usepackage[]{caption2}
\usepackage{graphicx}
\usepackage{blindtext}
\usepackage{color}
\usepackage{hyperref}
\usepackage{minted}
\usepackage{orcidlink}
\def\rmfcornisa{Research OR Education of Physics \hfill\rmf\ {\bf ??} (*?*) ???--???
\hfill MES? A\~NO?}

\def\rmfcintilla{{\it Rev.\ Mex.\ Fis.\/} {\bf ??} (*?*) (????) ???--???}
\clearpage \rmfcaptionstyle 
\begin{document}
\markboth{ RMF Editorial Team    }{ A \LaTeX template for the RMF, RMF-E, SRMF }

%
%
\title{Benchmarking a restricted Boltzmann machine on the $\mathbb{Z}_2$ Bose-Hubbard chain in the adiabatic hard-core regime
\vspace{-6pt}}
\author{Gustavo Alejandro Avalos Valentín\,\orcidlink{0000-0002-3739-4133}\thanks{E-mail:\href{mailto:regulus@ciencias.unam.mx}{\texttt{regulus@ciencias.unam.mx}}}}
\address{Instituto de Física, Universidad Nacional Autónoma de México,
Circuito de la Investigación Científica, Ciudad Universitaria, Ciudad de México 04510, México}
\author{ Roman Josué Armenta Rico\,\orcidlink{0000-0003-3004-025X}\thanks{E-mail: \href{mailto:romanarmenta@estudiantes.fisica.unam.mx}{\texttt{romanarmenta@estudiantes.fisica.unam.mx}}}}
\address{Instituto de Física, Universidad Nacional Autónoma de México,
Circuito de la Investigación Científica, Ciudad Universitaria, Ciudad de México 04510, México}
\author{Isaac P\'erez Castillo\,\orcidlink{0000-0001-7622-9440}\thanks{E-mail: \href{mailto:iperez@izt.uam.mx}{\texttt{iperez@izt.uam.mx}}}}
\address{Departamento de F\'isica, Universidad Aut\'onoma Metropolitana-Iztapalapa,
San Rafael Atlixco 186, Ciudad de M\'exico 09340, M\'exico}
\address{Instituto de Ciencias F\'isicas, Universidad Nacional Aut\'onoma de M\'exico (UNAM), Av. Universidad s/n, Col. Chamilpa, CP 62210 Cuernavaca, Mor., Mexico}

\maketitle

\begin{abstract}
\vspace{1em} 
We study the ground state of the $\mathbb{Z}_2$ Bose-Hubbard chain in the adiabatic hard-core limit at half filling using variational Monte Carlo with a shallow restricted Boltzmann machine as the variational ansatz. In this context, the neural quantum state is compared with the established adiabatic description of the model. The variational results reproduce the overall structure of the phase diagram obtained from magnetization observables, distinguish the polarized and N\'eel-ordered regions, and capture representative spin patterns and site occupations for the staggered insulating configurations selected by a weak symmetry-breaking field. Taken together, these results show that a shallow restricted Boltzmann machine reproduces the main adiabatic phase structure of the one-dimensional $\mathbb{Z}_2$ Bose-Hubbard chain and captures the selected symmetry-broken insulating configurations at half filling.
\vspace{1em}
\end{abstract}
\keys{ \bf{\textit{$\mathbb{Z}_2$ Bose-Hubbard model; restricted Boltzmann machine; neural quantum states; variational Monte Carlo; adiabatic hard-core limit.}} \vspace{-8pt}}
\begin{multicols}{2}

\section*{Introduction}
\label{sec:introduction}

The $\mathbb{Z}_2$ Bose-Hubbard model has emerged as a minimal one-dimensional strongly correlated lattice system in which interacting bosons are coupled to dynamical bond variables, providing a bosonic analog of Peierls physics where the role of the lattice distortion is played by discrete $\mathbb{Z}_2$ fields \cite{gonzalez2}. At half filling, and particularly in the adiabatic and hard-core limits, the model supports polarized phases and a N\'eel-ordered sector in which spontaneous symmetry breaking coexists with the insulating configurations identified in the literature as trivial and topological \cite{gonzalez2019symmetry}. Subsequent work considerably enriched this picture by analyzing emergent symmetry protection, fractionalized solitonic excitations, and extended Peierls-type structures in related regimes and generalizations of the model \cite{gonzalez2019intertwined,gonzalez2020solitons,chanda2022devil}. Recent reviews place these developments within the broader program of non-standard Bose-Hubbard models for quantum simulation with ultracold atoms \cite{chanda2025review}. These developments establish the $\mathbb{Z}_2$-Bose-Hubbard model as an active setting for the study of interacting bosons coupled to dynamical bond variables.

Beyond the original identification of a bosonic Peierls mechanism in a dynamical lattice and the half-filled symmetry-breaking topological insulator \cite{gonzalez2,gonzalez2019symmetry}, subsequent work clarified that this model belongs to a broader and still active line of research. In particular, emergent symmetry protection was shown to stabilize intertwined topological phases and to generate interaction-driven topological transitions together with fractional pumping phenomena \cite{gonzalez2019intertwined}. The defect sector was then extended beyond half filling in terms of $\mathbb{Z}_n$ solitons and boson fractionalization \cite{gonzalez2020zn,gonzalez2020solitons}, while related extensions exhibited Devil's-staircase structures of topological Peierls insulators and Peierls supersolids \cite{chanda2022devil}. More recently, the model has continued to appear in broader reviews of non-standard Bose-Hubbard systems for quantum simulation and in a DMRG study of the transverse-field dependence of its ground state \cite{chanda2025review,watanabe2025transverse}. 

From the numerical point of view, one-dimensional short-range many-body systems are still most naturally benchmarked against density-matrix renormalization group and related tensor-network methods, which remain the standard reference framework for ground-state calculations in this setting \cite{white,schollwock2011,catarina}. In parallel, neural quantum states have opened a different variational route to the many-body problem, beginning with Carleo and Troyer’s RBM-based neural-network representation of quantum many-body wave functions and subsequently expanding into a broad family of architectures, symmetry-preserving constructions, and optimization strategies \cite{carleo,melko,lange2024review}. Neural-network wave functions have also been explored in bosonic lattice settings, beginning with feedforward-network studies of the Bose-Hubbard model \cite{saito2017,saito2018}, continuing with RBM-based benchmarks of the one-dimensional Bose-Hubbard phase diagram \cite{mcbrian2019,vargas}, and extending to Bose-Hubbard ladders and specialized bosonic constructions tailored to the same model class \cite{ceven2022,pei2024}. In this context, we study the performance of a shallow RBM as a variational description of the $\mathbb{Z}_2$-Bose-Hubbard chain in the adiabatic hard-core regime.

The numerical context has also evolved substantially beyond the first generation of restricted-Boltzmann-machine applications. Recent reviews make clear that neural quantum states now comprise a broad family of architectures and applications \cite{lange2024review}. After the seminal RBM formulation \cite{carleo} and the early overview of RBMs in quantum physics \cite{melko}, the field expanded toward recurrent and autoregressive wave functions with exact sampling \cite{hibatallah2020rnn}, explicit applications to topological order beyond conventional Landau phases \cite{hibatallah2023topological}, and deeper analyses of expressive power linking deep neural quantum states to tensor-network constructions \cite{sharir2022tensor}. More recent progress has centered on large-scale optimization of deep neural quantum states \cite{chen2024deepnqs}, transformer-based variational ans\"atze \cite{sprague2024transformers}, Bose-Hubbard-specific specializations of neural quantum states that go beyond generic one-hot constructions \cite{pei2024}, and accurate bosonic lattice wave functions tailored to interacting Bose-Hubbard systems \cite{denis2025bosons}. At the same time, for one-dimensional short-range ground-state problems such as the present chain, DMRG and matrix-product-state methods remain the reference benchmark \cite{white,schollwock2011,catarina}.

In this work we consider the adiabatic limit $\beta = 0$, the hard-core boson limit $U\to\infty$, and half filling. Within this regime, the shallow RBM is used to reconstruct the gross phase structure from the total and staggered magnetizations of the $\mathbb{Z}_2$ fields and to examine representative symmetry-broken insulating configurations selected by a weak staggered field. The staggered field is introduced only to lift the degeneracy between the two Néel-ordered insulating configurations. The labels trivial and topological are used in the sense established in the literature \cite{gonzalez2019symmetry,gonzalez2019intertwined} for the two dimerization patterns, and not as the result of an independent invariant or edge-state calculation.

The remainder of the paper is organized as follows. In Sec. \ref{sec:z2-bose-hubbard-model-benchmark-regime}, we define the Hamiltonian and delimit the adiabatic hard-core regime at half filling that serves as the reference setting of the present benchmark. In Sec. \ref{sec:rbm-ansatz-variational-procedure}, we present the restricted Boltzmann machine ansatz, the variational optimization strategy, and the observables evaluated in this work. In Sec. \ref{sec:adiabatic-benchmark-magnetization-observables}, we discuss the phase structure reconstructed from the total and staggered magnetizations of the $\mathbb{Z}_2$ fields. In Sec. \ref{sec:selected-symmetry-broken-insulating-configurations}, we analyze representative symmetry-broken insulating configurations through their spin patterns and local boson occupations. In Sec. \ref{sec:scope-limitations-benchmark}, we make explicit the scope and limitations of the present benchmark. Finally, Sec. \ref{sec:conclusions} summarizes the main conclusions, and Appendix \ref{app:implementation-details} collects the implementation details relevant for reproducibility.

\section{The $\mathbb{Z}_2$ Bose-Hubbard model and the benchmark regime}
\label{sec:z2-bose-hubbard-model-benchmark-regime}

The model studied in this work is the one-dimensional $\mathbb{Z}_2$ Bose-Hubbard Hamiltonian
\begin{equation}
\begin{split}
\mathcal{H}_{\mathbb{Z}_2BH} &=-\alpha\sum_i\left(b_i^{\dagger}\sigma_{i,i+1}^{z}b_{i+1}+\mbox{H.c.}\right)+\beta\sum_i \sigma_{i,i+1}^{x}\\
&-t\sum_i\left(b_i^{\dagger}b_{i+1}+\mbox{H.c.}\right)+\frac{U}{2}\sum_i n_i(n_i-1)\\
&+\frac{\Delta}{2}\sum_i \sigma_{i,i+1}^{z}\,,
\end{split}
\label{eq:z2bh-hamiltonian}
\end{equation}
where $b_i^{\dagger}$ and $b_i$ are bosonic creation and annihilation operators on site $i$, $n_i=b_i^{\dagger}b_i$, and the Pauli matrices $\sigma_{i,i+1}^{x,z}$ act on the $\mathbb{Z}_2$ degree of freedom located on the link joining neighboring sites. The parameter $t$ denotes the bare boson tunneling, $\alpha$ controls the link-dependent modulation of that tunneling, $\beta$ is the transverse field that drives the dynamics of the $\mathbb{Z}_2$ sector, $U$ is the on-site repulsion, and $\Delta$ biases the two local link configurations. For a fixed background $\sigma_{i,i+1}^{z}=\pm1$, the hopping amplitudes become $-t\pm\alpha$. Although this Hamiltonian may be viewed as a bosonic $\mathbb{Z}_2$ lattice model, for the purposes of the present paper it is more convenient to regard the link spins as a minimal dynamical lattice that dresses the bosonic hopping and gives rise to a bosonic Peierls mechanism \cite{gonzalez2,gonzalez2019symmetry}.

\begin{figure}[H]
\centering
\includegraphics[scale=0.15]{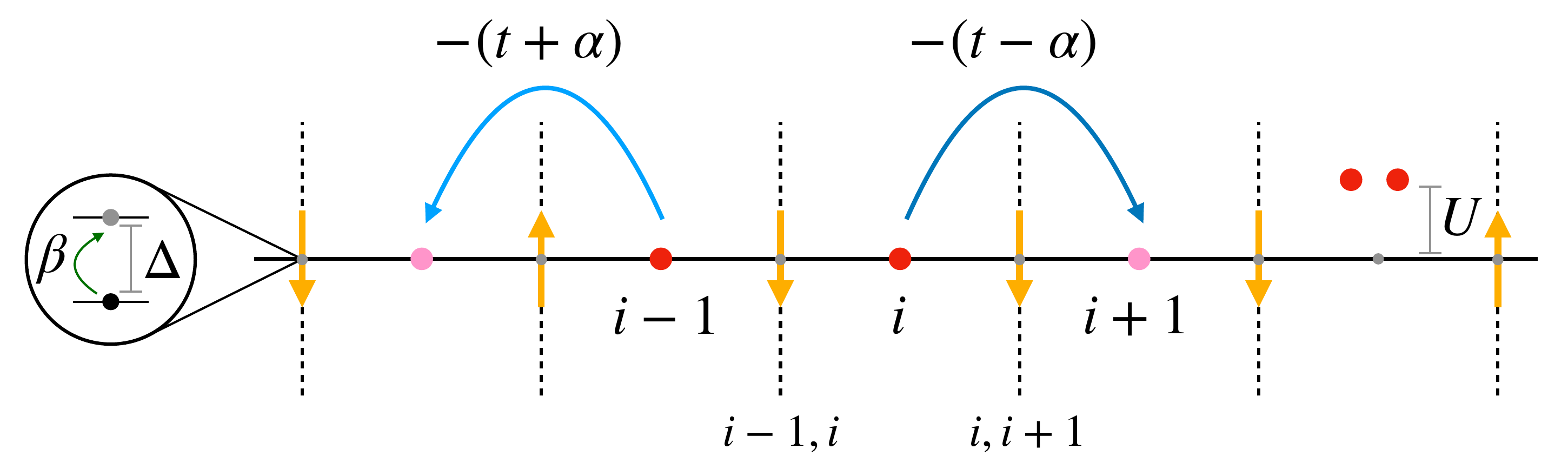}
\caption{Schematic representation of the one-dimensional $\mathbb{Z}_2$ Bose-Hubbard model. Bosons occupy the lattice sites and interact on-site with strength $U$, while the link variables $\sigma_{i,i+1}^{z}=\pm1$ modulate the hopping amplitudes into $-t\pm\alpha$. The parameters $\beta$ and $\Delta$ control, respectively, the transverse and longitudinal fields of the $\mathbb{Z}_2$ sector.}
\label{fig1}
\end{figure}

Throughout this manuscript we restrict attention to the benchmark regime that provides the cleanest test of the shallow RBM, namely half filling, the adiabatic limit $\beta=0$, and the hard-core constraint $U\rightarrow\infty$. At half filling the boson density is $\rho=N_b/L=1/2$, and in the hard-core limit double occupancy is forbidden. The bosons can then be fermionized by a Jordan-Wigner transformation, so that a static link background generates a dimerized hopping problem for spinless fermions \cite{gonzalez2019symmetry}. For the adiabatic reference solution it is convenient to divide the links into odd-even and even-odd sublattices, denoted by $A$ and $B$, and to characterize the $\mathbb{Z}_2$ background through the sublattice magnetizations $\langle\sigma_A^z\rangle$ and $\langle\sigma_B^z\rangle$. The two observables used later to benchmark the phase structure are the total and staggered magnetizations,
\begin{equation}
m_t=\frac{\langle\sigma_A^z\rangle+\langle\sigma_B^z\rangle}{2}\,,\qquad m_s=\frac{\langle\sigma_A^z\rangle-\langle\sigma_B^z\rangle}{2}\,,
\label{eq:z2bh-magnetizations}
\end{equation}
whose absolute values distinguish, respectively, the polarized sectors and the N\'eel-ordered sector.

In this limit the established reference picture is obtained from the Born-Oppenheimer treatment of the hard-core model \cite{gonzalez2019symmetry}. Because the spontaneous doubling of the unit cell reduces the problem to two inequivalent links, the $\mathbb{Z}_2$ background can be parametrized by two angles $\boldsymbol{\theta}=(\theta_A,\theta_B)$, and the corresponding variational energy reads
\begin{equation}
\begin{split}
\varepsilon_{\rm gs}(\boldsymbol{\theta})&=-\frac{2}{\pi}t(\boldsymbol{\theta})\,\mathrm{E}\left(1-\delta^2(\boldsymbol{\theta})\right)
+\frac{\Delta}{4}\left(\sin\theta_A+\sin\theta_B\right)\\
&-\frac{\beta}{2}\left(\cos\theta_A+\cos\theta_B\right)\,,
\label{eq:bo-energy}
\end{split}
\end{equation}
where
\begin{equation}
\begin{split}
t(\boldsymbol{\theta})&=t+\frac{\alpha}{2}\left(\sin\theta_A+\sin\theta_B\right)\,,\\
\delta(\boldsymbol{\theta})&= \frac{\alpha\left(\sin\theta_A-\sin\theta_B\right)}
{2t+\alpha\left(\sin\theta_A+\sin\theta_B\right)}\,,
\label{eq:bo-parameters}
\end{split}
\end{equation}
and $\mathrm{E}(x)=\int_{0}^{\pi/2}\mathrm{d}k\,\sqrt{1-x\sin^2k}$ is the complete elliptic integral of the second kind. For the adiabatic benchmark actually used in this paper, namely $\beta=0$, the minimization of Eq.~(\ref{eq:bo-energy}) yields the critical lines
\begin{equation}
\Delta_c^{\pm}=\frac{4t}{\pi}\left(\tilde{\delta}\pm \mathrm{E}\left(1-\tilde{\delta}^2\right)\mp 1\right)\,,\qquad\tilde{\delta}=\frac{\alpha}{t}\,,
\label{eq:adiabatic-critical-lines}
\end{equation}
which delimit the three regions that serve as the reference benchmark for the RBM calculations. In the remainder of the paper, this adiabatic hard-core solution is the only external reference used to assess the RBM phase maps.

The physical interpretation of Eq.~(\ref{eq:adiabatic-critical-lines}) is straightforward. For $\Delta>\Delta_c^+$ the minimum corresponds to the fully polarized configuration $|\downarrow\downarrow\downarrow\cdots\downarrow\rangle$, whereas for $\Delta<\Delta_c^-$ it corresponds to the fully polarized configuration $|\uparrow\uparrow\uparrow\cdots\uparrow\rangle$. In both cases translational invariance remains unbroken, $|m_t|=1$ and $m_s=0$ in the ideal ordered limit, and the effective hopping pattern is uniform. By contrast, for $\Delta\in(\Delta_c^-,\Delta_c^+)$ the minimum occurs at two symmetry-related N\'eel configurations, $|\uparrow\downarrow\uparrow\downarrow\cdots\uparrow\downarrow\rangle$ and $|\downarrow\uparrow\downarrow\uparrow\cdots\downarrow\uparrow\rangle$, for which $m_t=0$ and $|m_s|=1$ in the perfectly ordered limit. In the hard-core mapping these two backgrounds generate the two dimerizations of an SSH-type chain, and it is in this sense that the literature distinguishes a trivial and a topological insulating configuration \cite{ssh,gonzalez2019symmetry,gonzalez2019intertwined}. For hard-wall boundaries, the pattern $|\uparrow\downarrow\uparrow\downarrow\cdots\rangle$ corresponds to the trivial sector, whereas $|\downarrow\uparrow\downarrow\uparrow\cdots\rangle$ corresponds to the topological one \cite{gonzalez2019symmetry}.

Since the two N\'eel sectors are degenerate in the absence of an explicit selector, the representative finite-size calculations discussed later use a weak staggered field
\begin{equation}
\mathcal{H}_{\varepsilon}=\varepsilon\sum_i(-1)^i\sigma_{i,i+1}^{z},
\label{eq:staggered-selector}
\end{equation}
which lifts the degeneracy without changing the benchmark logic of the calculation. 
In the convention adopted below, $\varepsilon>0$ selects the configuration conventionally identified in the literature as trivial, while $\varepsilon<0$ selects the one conventionally identified as topological. This is the only sense in which those labels are used in the remainder of the paper: they refer to the established correspondence between the two symmetry-broken dimerization patterns and the effective SSH-type problem, not to an independent topological diagnosis carried out in the present manuscript.

\section{Restricted Boltzmann machine ansatz and variational procedure}
\label{sec:rbm-ansatz-variational-procedure}

To approximate the ground state of the model introduced in Sec.~\ref{sec:z2-bose-hubbard-model-benchmark-regime}, we employ a single-hidden-layer restricted Boltzmann machine (RBM), which is one of the standard neural-quantum-state ans\"atze introduced for variational many-body calculations \cite{carleo,melko}. Related neural-network-state constructions have also been explored in bosonic lattice models, including feedforward-network studies of the Bose-Hubbard model, RBM benchmarks of the one-dimensional Bose-Hubbard chain, Bose-Hubbard ladders, and more recent Bose-Hubbard-specific specializations of RBM-type ans\"atze \cite{saito2017,saito2018,mcbrian2019,vargas,ceven2022,pei2024}. For the finite open chains considered here, a basis state is written as
\begin{equation}
|\mathbf{n},\boldsymbol{\sigma}\rangle
=
|n_1,\ldots,n_L\rangle\otimes|\sigma_{1,2}^{z},\ldots,\sigma_{L-1,L}^{z}\rangle,
\label{eq:rbm-basis-state}
\end{equation}
where $n_i$ is the boson occupation on site $i$ and $\sigma_{i,i+1}^{z}=\pm1$ denotes the $\mathbb{Z}_2$ variable on the link $(i,i+1)$. The many-body state is then expanded as
\begin{equation}
|\Psi\rangle=\sum_{\{\mathbf{n}\},\{\boldsymbol{\sigma}\}}
\Psi(\mathbf{n},\boldsymbol{\sigma})\,|\mathbf{n},\boldsymbol{\sigma}\rangle.
\label{eq:rbm-many-body-state}
\end{equation}
Following the mixed bosonic-spin encoding used in RBM treatments of Bose-Hubbard-type models \cite{vargas}, each bosonic occupation is represented by one-hot visible variables
\begin{equation}
n_q^i=\delta_{q,n_i},
\qquad q=0,1,\ldots,n_{\mathrm{max}},
\label{eq:rbm-one-hot}
\end{equation}
while the link variables are kept as binary visible degrees of freedom. Here $q$ labels the occupation channel rather than a power: for a site with occupation $n_i$, the visible component $n_q^i$ equals one only for $q=n_i$ and vanishes otherwise. In the hard-core case, the local bosonic visible vector is therefore either (1,0) for an empty site or (0,1) for an occupied site. In the benchmark regime studied in the remainder of the paper, namely the hard-core limit, one has $n_{\mathrm{max}}=1$, so the bosonic sector reduces to two visible channels per site.

With this encoding, the visible layer of the RBM contains the set $\mathbf{v}=\bigl(\{n_q^i\},\{\sigma_{j,j+1}^{z}\}\bigr)$, and it is coupled to hidden Ising variables $\mathbf{h}=\{h_p\}_{p=1}^{M}$ with $h_p=\pm1$. The RBM energy function is taken as
\begin{equation}
\begin{split}
E_{\mathrm{RBM}}(\mathbf{v},\mathbf{h};\mathcal{W})&=-\sum_{j=1}^{L-1}a_j\,\sigma_{j,j+1}^{z}-\sum_{i=1}^{L}\sum_{q=0}^{n_{\mathrm{max}}} b_q^i\,n_q^i\\
&-\sum_{p=1}^{M}\sum_{j=1}^{L-1} A_{pj}\,h_p\,\sigma_{j,j+1}^{z}\\
&-\sum_{p=1}^{M}\sum_{i=1}^{L}\sum_{q=0}^{n_{\mathrm{max}}} B_{pq}^{i}\,h_p\,n_q^i
-\sum_{p=1}^{M} c_p\,h_p\,,
\end{split}
\label{eq:rbm-energy-function}
\end{equation}
where $\mathcal{W}=(\mathbf{a},\mathbf{b},\mathbf{c},A,B)$ collects the variational parameters. The corresponding RBM wave-function amplitude is obtained by summing over the hidden layer,
\begin{equation}
\psi_{\mathcal{W}}(\mathbf{n},\boldsymbol{\sigma}) = \sum_{\{\mathbf{h}\}} e^{-E_{\mathrm{RBM}}(\mathbf{v},\mathbf{h};\mathcal{W})}\,,
\label{eq:rbm-wavefunction-sum}
\end{equation}
which yields the explicit form
\begin{equation}
\begin{split}
\psi_{\mathcal{W}}(\mathbf{n},\boldsymbol{\sigma}) &= \exp\left(\sum_{j=1}^{L-1}a_j\,\sigma_{j,j+1}^{z}+\sum_{i=1}^{L}\sum_{q=0}^{n_{\mathrm{max}}} b_q^i\,n_q^i
\right)\\
&\hspace{-0.75cm}\times\prod_{p=1}^{M}2\cosh\left(\sum_{j=1}^{L-1} A_{pj}\,\sigma_{j,j+1}^{z}+\sum_{i=1}^{L}\sum_{q=0}^{n_{\mathrm{max}}}B_{pq}^{i}\,n_q^i+c_p\right)\,.
\label{eq:rbm-wavefunction}
\end{split}
\end{equation}
This is the shallow variational ansatz used throughout the present benchmark.

\begin{figure}[H]
\centering
\includegraphics[scale=0.17]{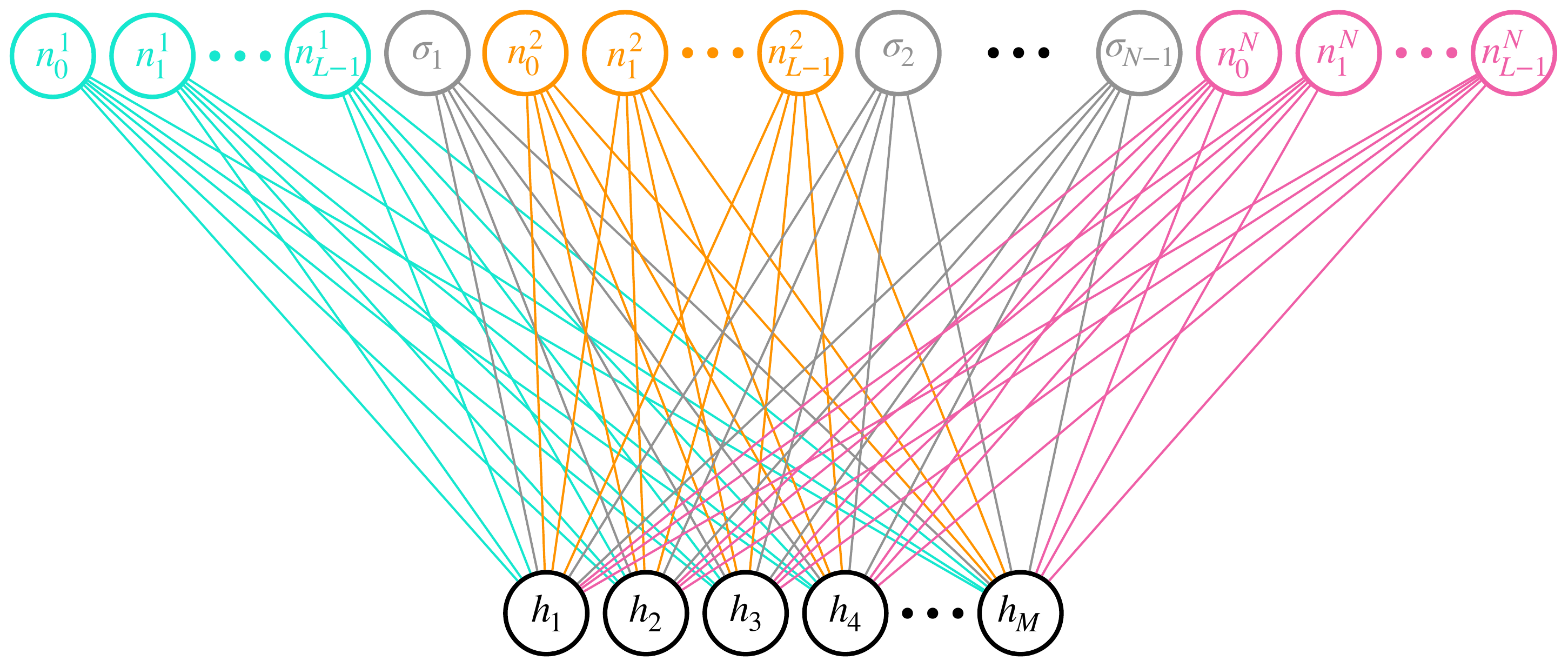}
\caption{Restricted Boltzmann machine used for the mixed bosonic-$\mathbb{Z}_2$ system. In the hard-core regime each bosonic site is represented by two one-hot visible units, $n_i^0$ and $n_i^1$, corresponding respectively to an empty or occupied site, while each link variable is represented by a binary visible unit $\sigma^z_{i,i+1}$. The hidden layer contains $M$ auxiliary neurons coupled to the visible layer through the parameters collected in $\mathcal{W}$.}
\label{fig2}
\end{figure}

The variational parameters are determined by minimizing the expectation value of the Hamiltonian. For notational brevity, in this section we denote by $\mathcal{H}$ either the bare Hamiltonian $\mathcal{H}_{\mathbb{Z}_2BH}$ or the symmetry-selected Hamiltonian $\mathcal{H}_{\mathbb{Z}_2BH}+\mathcal{H}_{\varepsilon}$ of Eq.~(\ref{eq:staggered-selector}), depending on the calculation under consideration. The local energy associated with a configuration $\mathcal{S}\equiv(\mathbf{n},\boldsymbol{\sigma})$ is
\begin{equation}
E_{\mathrm{loc}}(\mathcal{S})=\sum_{\mathcal{S}'}\langle\mathcal{S}|\mathcal{H}|\mathcal{S}'\rangle\frac{\psi_{\mathcal{W}}(\mathcal{S}')}{\psi_{\mathcal{W}}(\mathcal{S})}\,,
\label{eq:rbm-local-energy}
\end{equation}
and the corresponding variational energy is
\begin{equation}
\begin{split}
E_{\mathrm{var}}(\mathcal{W})&=\frac{\langle\psi_{\mathcal{W}}|\mathcal{H}|\psi_{\mathcal{W}}\rangle}
{\langle\psi_{\mathcal{W}}|\psi_{\mathcal{W}}\rangle}=\sum_{\mathcal{S}}P_{\mathcal{W}}(\mathcal{S})\,E_{\mathrm{loc}}(\mathcal{S})\,,\\
P_{\mathcal{W}}(\mathcal{S})&=\frac{|\psi_{\mathcal{W}}(\mathcal{S})|^{2}}
{\sum_{\mathcal{S}'}|\psi_{\mathcal{W}}(\mathcal{S}')|^{2}}\,.
\label{eq:rbm-variational-energy}
\end{split}
\end{equation}
The Monte Carlo procedure therefore samples configurations according to the probability distribution $P_{\mathcal{W}}(\mathcal{S})$ and estimates the energy from the average of the local estimator in Eq.~(\ref{eq:rbm-local-energy}).

In practice, the sampling is carried out with local Metropolis updates in the mixed bosonic-spin Hilbert space as implemented in NetKet \cite{netket3:2021}. Starting from an admissible configuration in the chosen particle-number sector, a candidate configuration $\mathcal{S}'$ is proposed through a local update and is accepted with probability
\begin{equation}
\mathcal{P}_{\mathrm{acc}}(\mathcal{S}\rightarrow\mathcal{S}')=\min\left\{1,\left|\frac{\psi_{\mathcal{W}}(\mathcal{S}')}{\psi_{\mathcal{W}}(\mathcal{S})}\right|^{2}\right\}\,.
\label{eq:rbm-acceptance}
\end{equation}
After thermalization, the resulting Markov chain provides the configurations used to estimate Eq.~(\ref{eq:rbm-variational-energy}) and the expectation values of the observables. The parameters $\mathcal{W}$ are then updated iteratively so as to decrease the variational energy. Since the explicit update equations are not specific to the $\mathbb{Z}_2$ Bose-Hubbard problem, we defer the generic formulas for gradient-based optimization and stochastic reconfiguration \cite{sorella} to Appendix \ref{app:implementation-details}.

Once the optimization has converged, all observables reported in the next sections are evaluated as Monte Carlo averages over the optimized distribution $P_{\mathcal{W}}$. In the phase-diagram benchmark of Sec.~\ref{sec:adiabatic-benchmark-magnetization-observables}, the relevant quantities are the total and staggered magnetizations defined in Eq.~(\ref{eq:z2bh-magnetizations}). In the representative symmetry-broken configurations discussed in Sec.~\ref{sec:selected-symmetry-broken-insulating-configurations}, we additionally compute the link-resolved profile $\langle\sigma_{i,i+1}^{z}\rangle$ and the local boson occupations $\langle n_i\rangle$.

\section{Adiabatic benchmark from magnetization observables}
\label{sec:adiabatic-benchmark-magnetization-observables}

\begin{figure}[H]
\centering
    \includegraphics[scale=0.25]{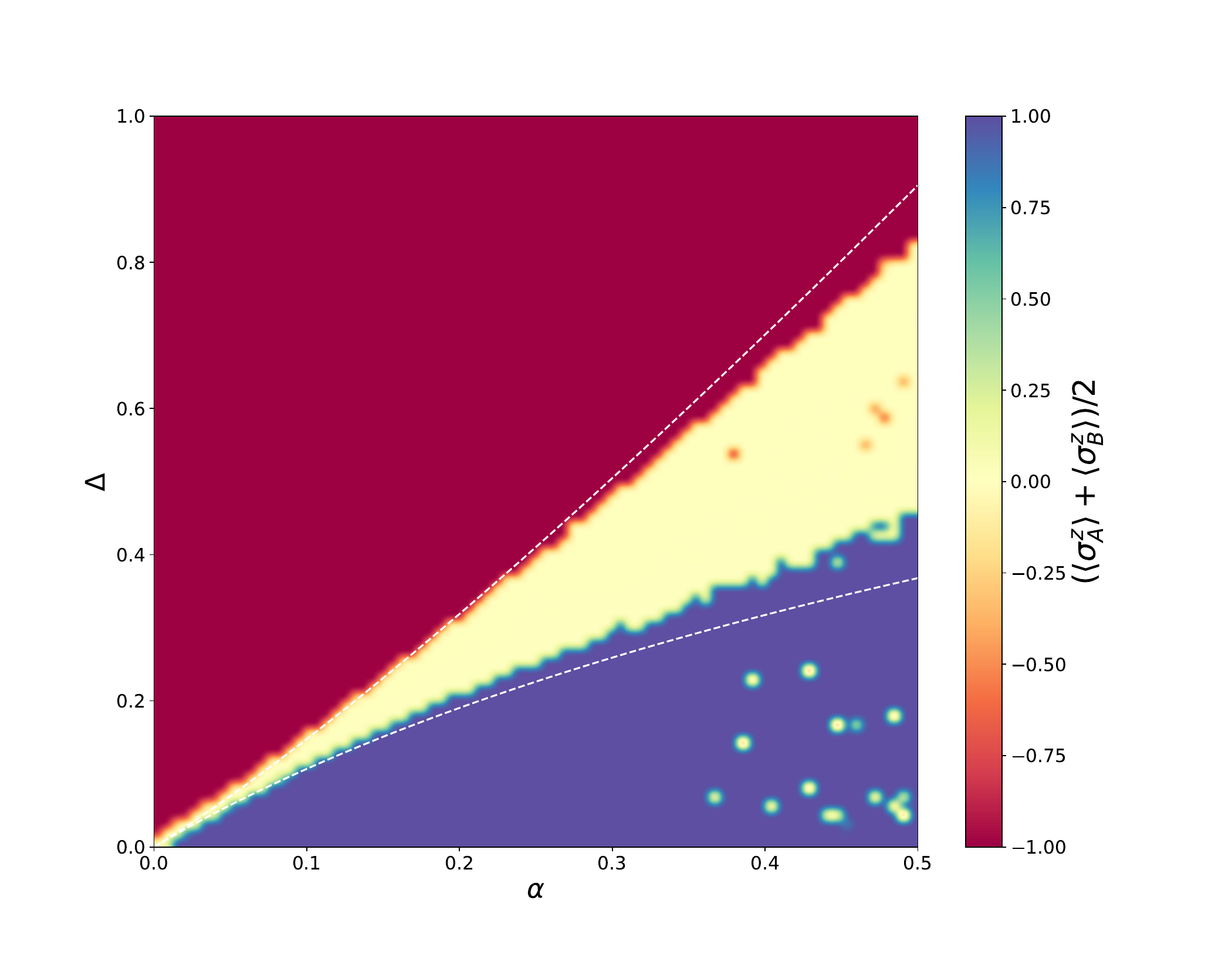}
    \caption{RBM estimate of the total magnetization \(m_t=(\langle\sigma_A^z\rangle+\langle\sigma_B^z\rangle)/2\) in the adiabatic hard-core benchmark. The lower and upper regions correspond to the two polarized backgrounds, while the intermediate region with \(m_t\approx 0\) identifies the loss of net polarization. The dashed white lines show the critical lines of Eq.~\ref{eq:adiabatic-critical-lines}.}
    \label{fig3}
\end{figure}

\begin{figure}[H]
\centering
    \includegraphics[scale=0.25]{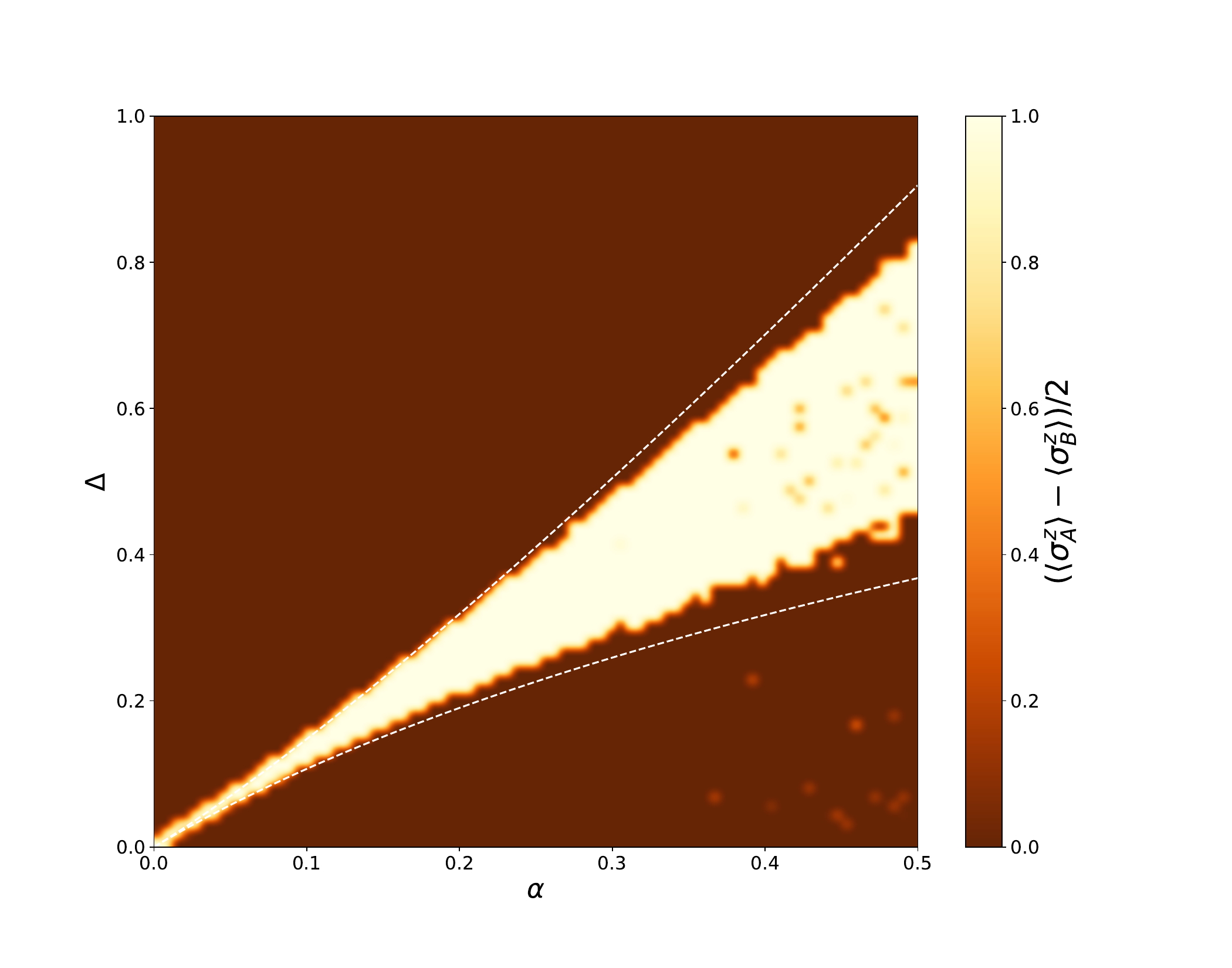}
    \caption{RBM estimate of the absolute staggered magnetization \(|m_s|=|\langle\sigma_A^z\rangle-\langle\sigma_B^z\rangle|/2\) for the same parameter plane as Fig. \ref{fig3}. The intermediate wedge is the region where the staggered response becomes maximal, confirming the N\'eel-ordered sector. The dashed white lines again show the critical lines of Eq. \ref{eq:adiabatic-critical-lines}.}
    \label{fig4}
\end{figure}

The first test of the shallow RBM is whether it reproduces the order-parameter structure of the established adiabatic reference solution. To this end, we fix the energy scale by setting \(t=1\), restrict to the adiabatic hard-core regime \(\beta=0\) and \(U\rightarrow\infty\) at half filling, and optimize an independent RBM state at each point of the \((\alpha,\Delta)\) grid shown in Figs. \ref{fig3} and \ref{fig4}. The optimized state is then probed through the total and staggered magnetizations introduced in Eq.~\eqref{eq:z2bh-magnetizations}. Since the two symmetry-related N\'eel configurations differ only by the sign of the staggered response, the phase map is most conveniently summarized by the pair \((m_t,|m_s|)\).
 We first examine whether these observables reproduce the adiabatic critical structure encoded in Eq.~\ref{eq:adiabatic-critical-lines} and in the Born-Oppenheimer analysis of Ref. \cite{gonzalez2019symmetry}.

Figure \ref{fig3} displays the total magnetization \(m_t\). The RBM results recover the expected three-region structure of the adiabatic hard-core phase diagram. For small \(\Delta\), the optimized states lie in a polarized sector with \(m_t\approx +1\), whereas for sufficiently large \(\Delta\) they lie in the opposite polarized sector with \(m_t\approx -1\). Between these two regions there appears an intermediate wedge where the net magnetization is strongly suppressed, \(m_t\approx 0\), which is precisely the behavior expected when the translationally invariant backgrounds give way to a doubled-unit-cell configuration. The broadening of this intermediate wedge as \(\alpha\) increases follows the same overall trend as the adiabatic critical lines, and the closing of the wedge near \(\alpha=0\) is also captured by the RBM map. At the level of overall phase structure, Fig. \ref{fig3} therefore shows that the shallow ansatz separates the two polarized sectors from the intermediate N\'eel-ordered sector in the correct region of parameter space.

The interpretation of the intermediate region is confirmed by the staggered response. Figure \ref{fig4} shows that the same wedge singled out by \(m_t\approx 0\) is the region where \(|m_s|\) becomes large, whereas both polarized sectors remain close to \(|m_s|=0\). The combined reading of Figs. \ref{fig3} and \ref{fig4} is therefore fully consistent with the adiabatic reference picture: outside the wedge the \(\mathbb{Z}_2\) background is polarized and translational symmetry remains unbroken, while inside it the link variables develop N\'eel order and the unit cell doubles, in agreement with the Born-Oppenheimer description of the half-filled hard-core chain \cite{gonzalez2019symmetry}. The magnetizations alone identify the common symmetry-broken sector, but they do not distinguish between the two degenerate dimerization patterns that the literature labels as trivial and topological. That distinction is addressed only in the next section by adding a weak staggered selector field.

The broadened interfaces and isolated outliers visible in Figs. \ref{fig3} and \ref{fig4} indicate that the present RBM data should be interpreted as a reconstruction of the overall adiabatic phase structure rather than as a high-precision determination of the transition lines. The outliers may reflect initialization effects, finite sampling, optimizer limitations, or limitations of the shallow ansatz; distinguishing among these possibilities would require a dedicated multi-start and architecture-scaling study. The phase boundaries should therefore be read primarily from the established adiabatic reference solution, while the RBM data support the qualitative separation between polarized and N\'eel-ordered regions.

\section{Selected symmetry-broken insulating configurations}
\label{sec:selected-symmetry-broken-insulating-configurations}

The magnetization maps discussed in Sec.~\ref{sec:adiabatic-benchmark-magnetization-observables} identify the symmetry-broken sector only through the common N\'eel order of the $\mathbb{Z}_2$ links. To inspect the two degenerate dimerization patterns separately, we now add the weak selector field of Eq.~\eqref{eq:staggered-selector} and analyze the representative finite-size states already contained in the present calculations. For the open chain considered here, the two signs of the selector field pick the two N\'eel backgrounds that are conventionally associated in the literature with the trivial and topological insulating sectors of the half-filled hard-core model \cite{gonzalez2019symmetry,gonzalez2019intertwined}. We stress again that these labels are used here only in that literature-based sense: the present data distinguish between the two selected symmetry-broken configurations, but they do not establish topology through an independent invariant, edge-state analysis, or entanglement diagnostic.

For the representative plots in this section, we keep the same adiabatic hard-core parameters used in the original calculations, namely \(t=1\), \(\alpha=0.5\), \(\Delta=1\), \(\beta=0\), and \(U\rightarrow\infty\), and we lift the degeneracy with a weak staggered field of magnitude \(|\varepsilon|=0.05\). The sign of \(\varepsilon\) determines which of the two symmetry-related N\'eel configurations is selected by the RBM optimization.

\begin{figure}[H]
    \centering
    \includegraphics[scale=0.18]{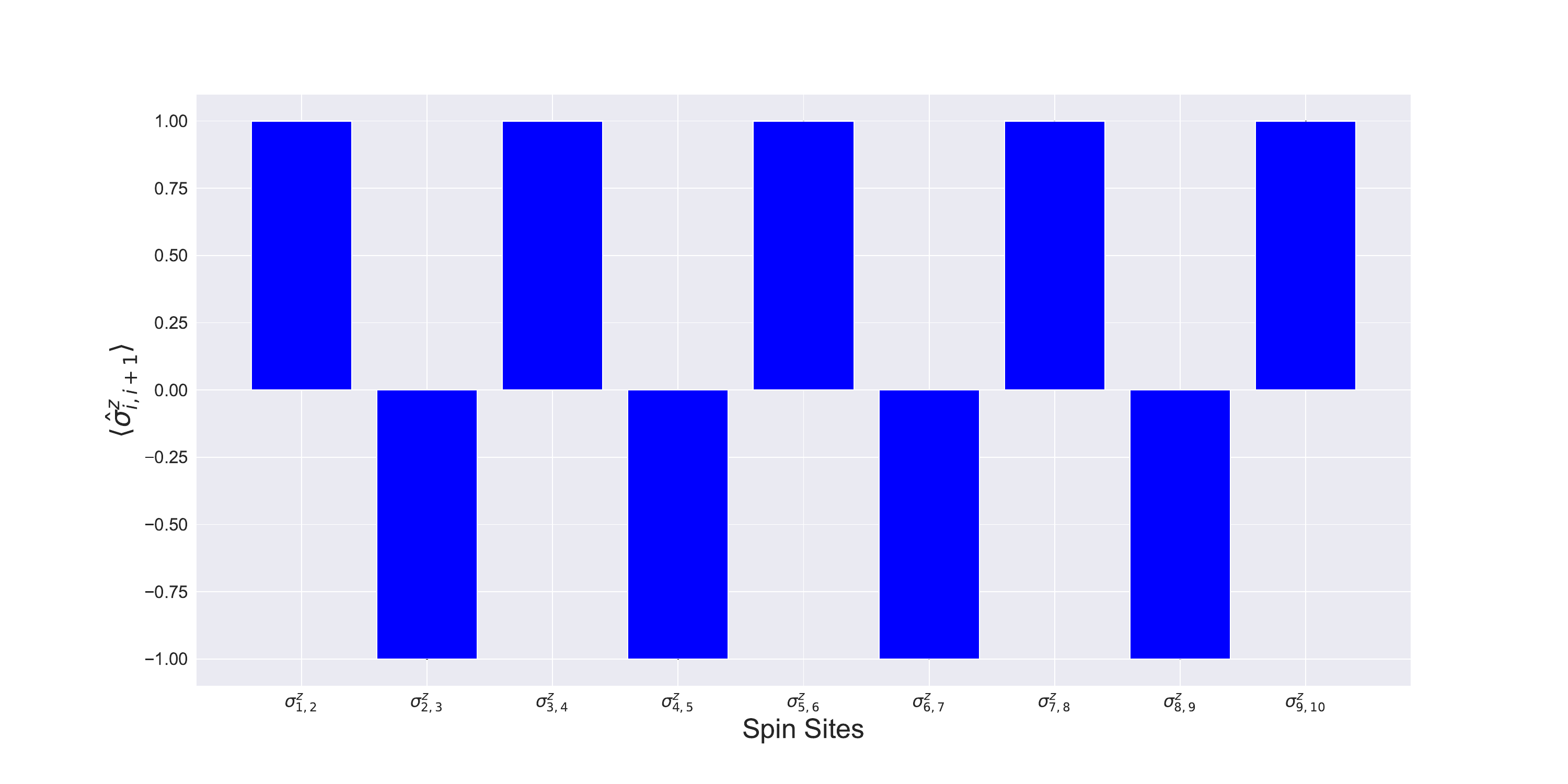}
    \caption{RBM estimate of the link expectation values $\langle \sigma^z_{i,i+1}\rangle$ for the selector-field choice $\varepsilon>0$. The staggered field lifts the N\'eel degeneracy and selects the configuration conventionally associated with the trivial insulating sector.}
    \label{fig5}
\end{figure}

When $\varepsilon>0$, the optimized state converges to the staggered link pattern shown in Fig.~\ref{fig5}. The expectation values $\langle \sigma^z_{i,i+1}\rangle$ alternate with essentially saturated sign from link to link, showing that the optimized state follows the selected broken-symmetry background once the external selector removes the degeneracy. In the convention adopted here, this is the configuration associated in the literature with the trivial insulating sector. At the level of the Monte Carlo uncertainty shown in the figure, the optimized state displays the selected alternating link background at this representative point.

\begin{figure}[H]
    \centering
    \includegraphics[scale=0.18]{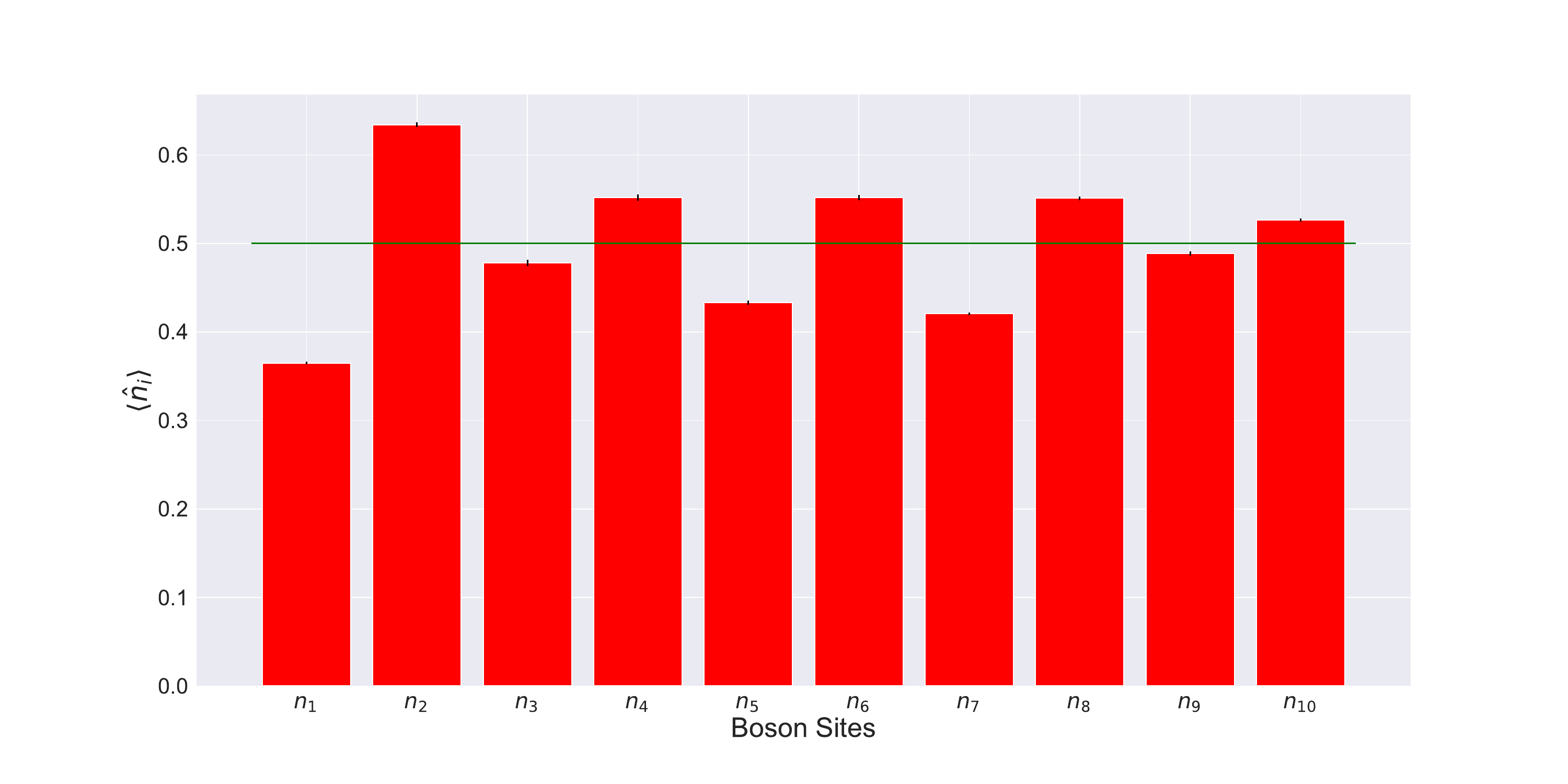}
    \caption{RBM estimate of the local boson occupations $\langle n_i\rangle$ for the same representative point as in Fig.~\ref{fig5}. The green line marks the target half filling.}
    \label{fig6}
\end{figure}

The corresponding bosonic density profile is displayed in Fig.~\ref{fig6}. Its spatial average remains consistent with the target half filling, while the site-resolved occupations show a visible modulation around that average. Thus, for the $\varepsilon>0$ representative state, the selected configuration is identified most directly through the alternating link order, while the density profile provides a consistency check of the half-filled bosonic sector.

\begin{figure}[H]
    \centering
    \includegraphics[scale=0.18]{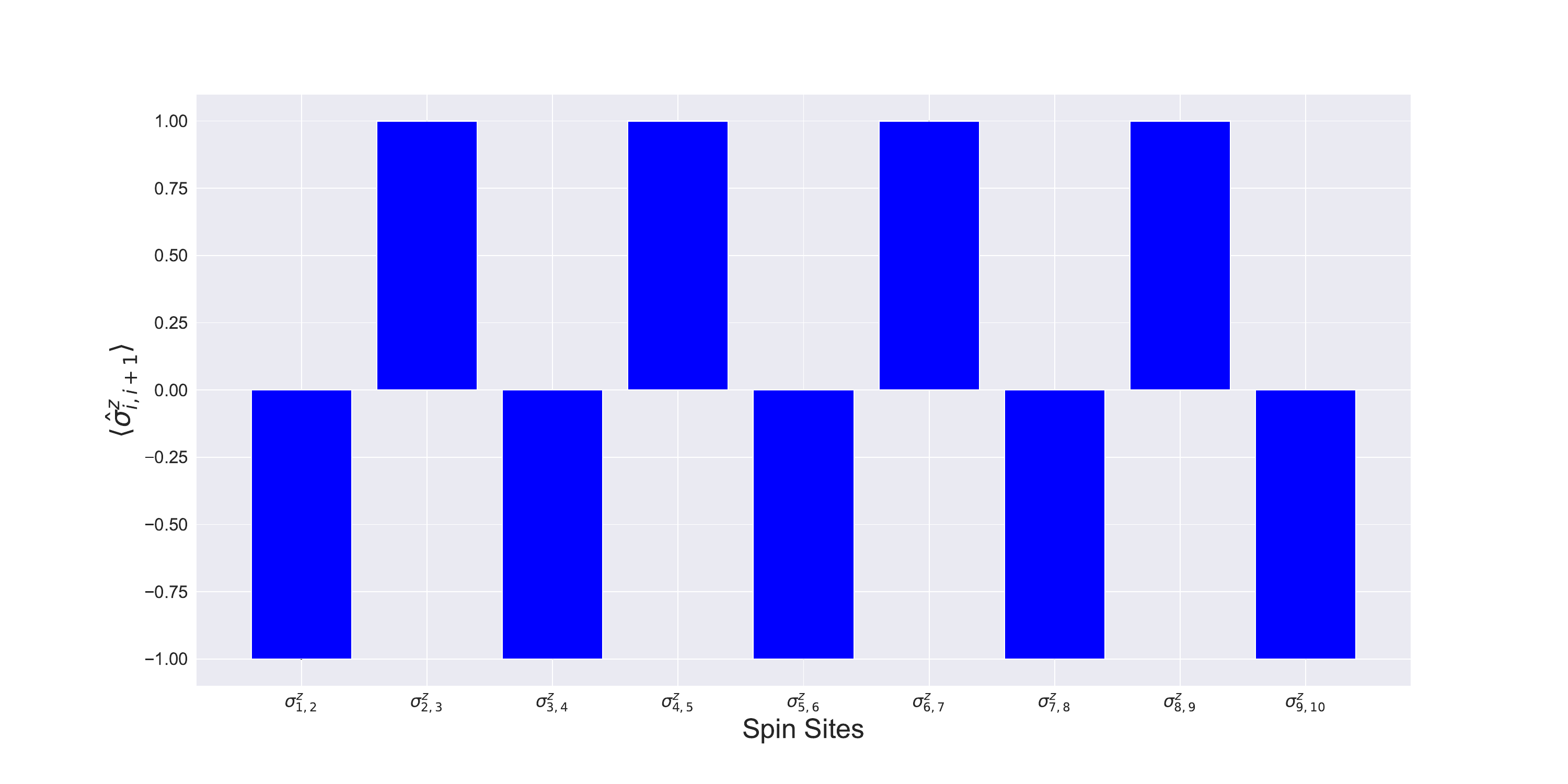}
    \caption{RBM estimate of the link expectation values $\langle \sigma^z_{i,i+1}\rangle$ for the selector-field choice \(\varepsilon<0\). In the convention adopted here, this is the symmetry-broken configuration conventionally associated with the topological insulating sector.}
    \label{fig7}
\end{figure}

Reversing the sign of the selector field exchanges the two symmetry-related dimerizations. For $\varepsilon<0$, the optimized RBM state converges to the reversed N\'eel ordering shown in Fig.~\ref{fig7}, where the sign of every link expectation value is inverted with respect to Fig.~\ref{fig5}. In the literature of the half-filled hard-core chain, this is the configuration conventionally associated with the topological insulating sector \cite{gonzalez2019symmetry,gonzalez2019intertwined}. This shows that the same shallow ansatz also represents the second broken-symmetry configuration once the external field fixes the sign of the dimerization. As in the previous case, the error bars are too small to affect the qualitative reading of the figure.

\begin{figure}[H]
    \centering
    \includegraphics[scale=0.18]{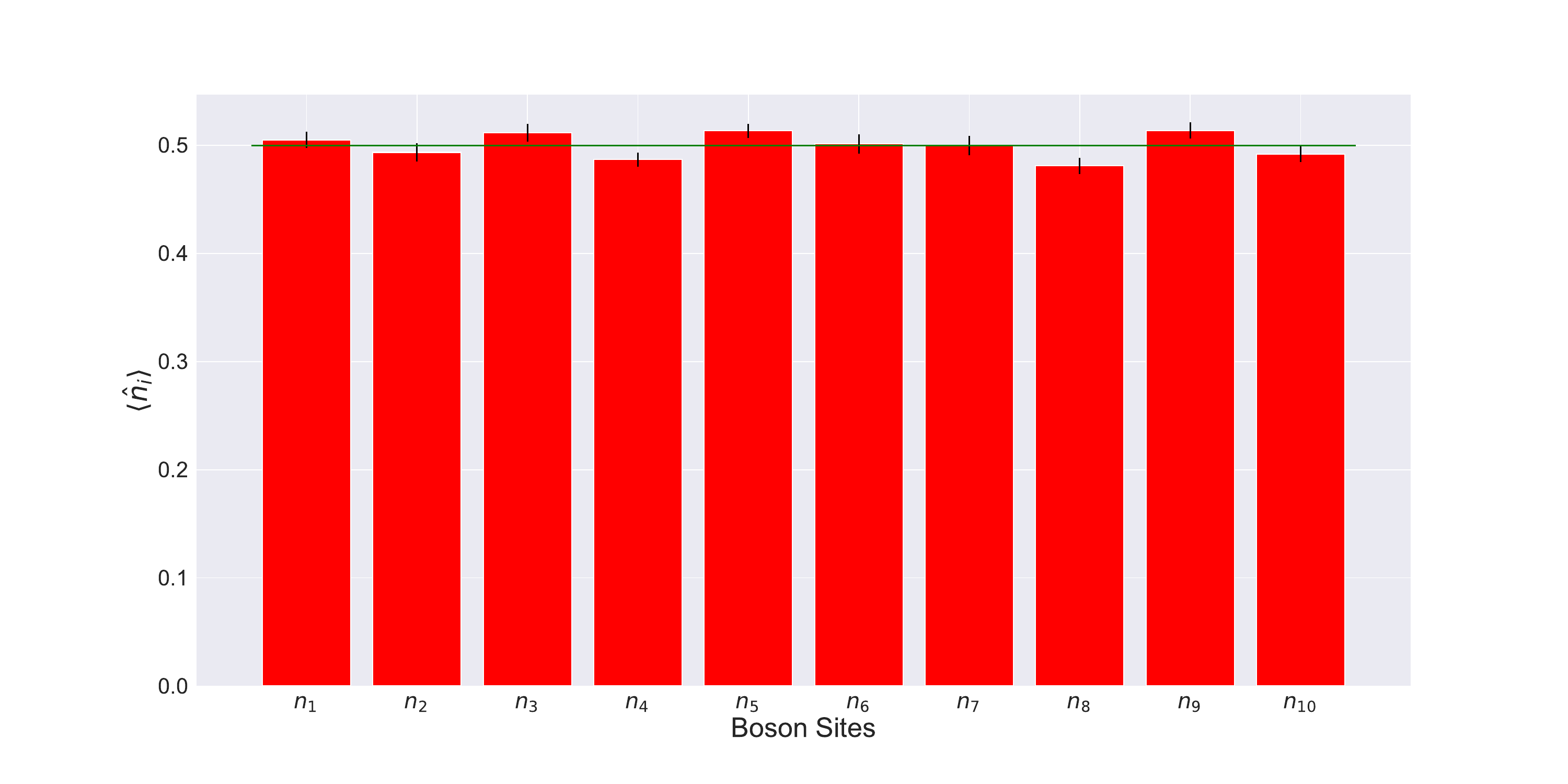}
    \caption{RBM estimate of the local boson occupations $\langle n_i\rangle$ for the same representative point as in Fig.~\ref{fig7}. The green line marks the target half filling.}
    \label{fig8}
\end{figure}

The density profile corresponding to this second representative state is shown in Fig.~\ref{fig8}. As in the $\varepsilon>0$ case, the spatial average is consistent with half filling. The site-resolved occupations appear less strongly modulated around the average in this representative run, but the essential diagnostic remains the reversal of the staggered link pattern shown in Fig.~\ref{fig7}. Together, Figs.~\ref{fig7} and \ref{fig8} show that the optimized RBM state represents the selected reversed dimerization while maintaining the bosonic sector close to half filling.

Taken together, Figs.~\ref{fig5}--\ref{fig8} show that the shallow RBM can represent both symmetry-broken insulating configurations selected from the adiabatic hard-core N\'eel sector. The optimized states reproduce the alternating link patterns and keep the bosonic density close to half filling in both selector sectors. The distinction between the trivial and topological cases is inherited from the established identification of the two dimerizations in the underlying SSH-type mapping, not from an independent topological diagnostic performed here.

\section{Scope and limitations of the benchmark}
\label{sec:scope-limitations-benchmark}
For the one-dimensional $\mathbb{Z}_2$ Bose-Hubbard chain at half filling, in the adiabatic limit $\beta=0$ and the hard-core limit $U\rightarrow\infty$, a shallow RBM reproduces the gross order-parameter structure of the known reference phase diagram and represents the two symmetry-broken insulating configurations once a weak staggered field removes their degeneracy.  The results reported here are restricted to this regime and to the total and staggered magnetizations of Eq.~(\ref{eq:z2bh-magnetizations}), the link-resolved expectation values $\langle \sigma_{i,i+1}^{z}\rangle$, and the local boson occupations $\langle n_i\rangle$. 

A first limitation concerns the interpretation of the two dimerized insulating configurations. In the hard-core mapping, the two N\'eel backgrounds correspond to the two dimerizations of an SSH-type chain and are conventionally identified in the literature as trivial and topological sectors \cite{ssh,gonzalez2019symmetry,gonzalez2019intertwined}. In the present manuscript, however, these labels are inherited from that established correspondence. We do not compute edge states, many-body polarization, Berry or Zak phases, entanglement-spectrum degeneracies, or other nonlocal indicators that would provide an independent topological diagnosis. The two cases analyzed in Sec.~\ref{sec:selected-symmetry-broken-insulating-configurations} should therefore be understood strictly as externally selected symmetry-broken configurations, together with the literature-based identification of those configurations.

A second qualification concerns the numerical scope of the ansatz itself. The broadened interfaces and isolated outliers in the magnetization maps show that the present RBM data recover the organization of the adiabatic phases, but they are not intended to replace the Born-Oppenheimer reference solution as a high-precision determination of the transition lines \cite{gonzalez2019symmetry}. More generally, for one-dimensional short-range ground-state problems, density-matrix renormalization group and matrix-product-state methods remain the standard reference methods \cite{white,schollwock2011,catarina}, while the current neural-quantum-state landscape includes architectures and optimization strategies substantially more expressive than a single-hidden-layer RBM, as well as bosonic benchmarks and Bose-Hubbard-specific constructions that go beyond generic one-hot encodings \cite{lange2024review,ceven2022,pei2024,chen2024deepnqs,sprague2024transformers,denis2025bosons}. The claim made here is therefore deliberately focused: within the controlled adiabatic hard-core regime, a shallow RBM already captures the main order-parameter structure and the selected symmetry-broken configurations.

A third qualification concerns regimes beyond the adiabatic hard-core setting. Away from $\beta=0$ and $U\rightarrow\infty$, the $\mathbb{Z}_2$ Bose-Hubbard model exhibits a richer phase structure involving nonadiabatic and finite-interaction effects \cite{gonzalez2019symmetry,chanda2025review,watanabe2025transverse}. Extending the present RBM analysis to that regime would require additional observables, such as superfluid stiffness or compressibility, together with a systematic comparison to established reference methods. We therefore restrict the physical conclusions of this work to the adiabatic hard-core regime analyzed above.

Natural extensions of this work include multi-start optimization at the outlier points, systematic architecture scaling, finite-size analysis, comparison with tensor-network data, and independent topological diagnostics such as edge-state analysis, many-body polarization, or entanglement-based probes.

Taken together, these points delimit the regime addressed by the present results. In the adiabatic hard-core limit at half filling, the shallow RBM reproduces the main phase structure of the known reference solution and represents the two symmetry-broken insulating configurations selected by a weak staggered field. The full nonadiabatic phase diagram and an independent topological characterization remain outside the scope of the present calculations.

\section{Conclusions}
\label{sec:conclusions}
In this work we studied the one-dimensional $\mathbb{Z}_2$ Bose-Hubbard chain in the adiabatic hard-core regime at half filling using variational Monte Carlo with a shallow restricted Boltzmann machine as the variational ansatz. Within this scope, the variational calculations reproduce the gross phase structure of the established reference solution \cite{gonzalez2019symmetry}. In particular, the RBM distinguishes the two polarized regions and the intermediate N\'eel-ordered region through the total and staggered magnetizations, so at the level of order-parameter structure it captures the main organization of the adiabatic phase diagram.

A second conclusion concerns representative states inside the symmetry-broken sector. By introducing a weak staggered field that lifts the finite-size degeneracy between the two dimerized backgrounds, the RBM converges to both selected configurations and reproduces their alternating link patterns and local boson occupations around half filling. In the terminology adopted from the literature, these are the configurations conventionally associated with the trivial and topological insulating sectors \cite{gonzalez2019symmetry,gonzalez2019intertwined}. The present results therefore show that the shallow ansatz can approximate both selected dimerizations, but they do not by themselves establish topology through an independent nonlocal diagnostic.

Taken together, these results show that a shallow RBM captures the main adiabatic symmetry structure of the $\mathbb{Z}_2$ Bose-Hubbard chain and represents the selected symmetry-broken insulating configurations in the regime studied here. The present calculations do not determine the phase boundaries with high precision, do not provide an independent topological characterization, and do not validate the quasi-adiabatic regime.

\appendix
\section{Implementation details}
\label{app:implementation-details}

This appendix collects the optimization formulas and numerical settings directly used in the present benchmark. For a configuration $\mathcal{S}\equiv(\mathbf{n},\boldsymbol{\sigma})$, the RBM state defines the sampling distribution
\begin{equation}
P_{\mathcal{W}}(\mathcal{S})=\frac{|\psi_{\mathcal{W}}(\mathcal{S})|^2}{\sum_{\mathcal{S}'}|\psi_{\mathcal{W}}(\mathcal{S}')|^2}\,.
\label{eq:app-probability-distribution}
\end{equation}
Accordingly, the expectation value of a generic observable \(\hat{O}\) can be written as
\begin{equation}
\langle \hat{O}\rangle=\sum_{\mathcal{S}}P_{\mathcal{W}}(\mathcal{S})\,\tilde{O}(\mathcal{S})\,,\qquad
\tilde{O}(\mathcal{S})=\frac{\langle \mathcal{S}|\hat{O}|\psi_{\mathcal{W}}\rangle}
{\langle \mathcal{S}|\psi_{\mathcal{W}}\rangle}\,,
\label{eq:app-local-estimator}
\end{equation}
and is estimated in practice from a Monte Carlo sample $\{\mathcal{S}_m\}_{m=1}^{\mathcal{N}}$ as
\begin{equation}
\langle \hat{O}\rangle\approx\frac{1}{\mathcal{N}}\sum_{m=1}^{\mathcal{N}}\tilde{O}(\mathcal{S}_m)\,.
\label{eq:app-monte-carlo-average}
\end{equation}
For $\hat{O}=\mathcal{H}$, the estimator $\tilde{O}$ reduces to the local energy already introduced in Eq.~\eqref{eq:rbm-local-energy}, and Eq.~\eqref{eq:app-monte-carlo-average} gives the Monte Carlo estimate of the variational energy in Eq.~\eqref{eq:rbm-variational-energy}.

The derivatives entering the optimization are conveniently expressed in terms of the logarithmic derivatives of the variational state,
\begin{equation}
O_i(\mathcal{S})\equiv\frac{\partial}{\partial \mathcal{W}_i}\ln \psi_{\mathcal{W}}(\mathcal{S})\,,
\label{eq:app-log-derivative}
\end{equation}
where $\mathcal{W}_i$ denotes any component of the RBM parameter set $\mathcal{W}=(\mathbf{a},\mathbf{b},\mathbf{c},A,B)$. Using Eq.~\eqref{eq:app-log-derivative}, the energy gradient can be written as
\begin{equation}
g_i\equiv\frac{\partial E_{\mathrm{var}}}{\partial \mathcal{W}_i}=
\left\langle E_{\mathrm{loc}}\,O_i^{*}\right\rangle-\left\langle E_{\mathrm{loc}}\right\rangle\left\langle O_i^{*}\right\rangle=
\left\langle\left(E_{\mathrm{loc}}-\langle E_{\mathrm{loc}}\rangle\right)O_i^{*}
\right\rangle \,.
\label{eq:app-energy-gradient}
\end{equation}
A simple gradient-descent update therefore reads
\begin{equation}
\mathcal{W}_i^{(p+1)}= \mathcal{W}_i^{(p)}-\eta\, g_i^{(p)}\,,
\label{eq:app-gradient-descent}
\end{equation}
with $\eta$ the learning rate and $p$ the optimization step.

Besides plain gradient descent, the original implementation also considered stochastic reconfiguration (SR), which incorporates the local geometry of the variational manifold \cite{sorella}. In SR, the parameter increment $\delta\mathcal{W}_i$ is obtained from the linear system
\begin{equation}
\sum_j S_{ij}\delta\mathcal{W}_j = -\gamma f_i\,,
\label{eq:app-sr-linear-system}
\end{equation}
followed by the update
\begin{equation}
\mathcal{W}_i^{(p+1)}= \mathcal{W}_i^{(p)}+\delta\mathcal{W}_i \,.
\label{eq:app-sr-update}
\end{equation}
Here $\gamma$ is the SR step size, $S_{ij}$ is the covariance matrix of the logarithmic derivatives,
\begin{equation}
S_{ij}=\left\langle O_i^{*}O_j\right\rangle-\left\langle O_i^{*}\right\rangle\left\langle O_j\right\rangle\,,
\label{eq:app-sr-metric}
\end{equation}
and $f_i$ are the generalized forces,
\begin{equation}
f_i=\left\langle E_{\mathrm{loc}}\,O_i^{*}\right\rangle-\left\langle E_{\mathrm{loc}}\right\rangle\left\langle O_i^{*}\right\rangle\,.
\label{eq:app-sr-forces}
\end{equation}
In practice, the covariance matrix can become ill-conditioned, so the SR step is implemented through a pseudo-inverse or a regularized inverse of \(S\), as is standard in variational Monte Carlo calculations \cite{sorella}.

The Markov chain used to sample Eq.~\eqref{eq:app-probability-distribution} was generated with NetKet's local Metropolis sampler \cite{netket3:2021}. In the fixed-half-filling calculations, the Markov chain was restricted to the chosen particle-number sector. Bosonic proposals were therefore required to preserve the total number of hard-core bosons, or were rejected whenever they would leave the half-filled sector. Updates of the $\mathbb{Z}_2$ variables changed the corresponding link state. The move was then accepted with the probability given in Eq.~\eqref{eq:rbm-acceptance}. In the hard-core benchmark reported in the main text, $n_{\mathrm{max}}=1$, so the bosonic visible layer reduces to two channels per site. As in the original calculations, the Monte Carlo sample entering each variational estimate was built from $1200$ local Metropolis iterations, and the sampling and parameter updates were repeated until convergence of the RBM parameters.

For the adiabatic benchmark of Sec.~\ref{sec:adiabatic-benchmark-magnetization-observables}, the parameter plane was explored on a $40\times40$ grid in $(\alpha,\Delta)$ with $t=1$, $\beta=0$, $U\rightarrow\infty$, and half filling, performing an independent RBM optimization at each grid point. For the representative symmetry-broken configurations of Sec.~\ref{sec:selected-symmetry-broken-insulating-configurations}, the parameters were fixed to $t=1$, $\alpha=0.5$, $\Delta=1$, $\beta=0$, and $U\rightarrow\infty$, while a weak selector field with $|\varepsilon|=0.05$ was used to isolate the two N\'eel sectors. In those representative runs, the evolution of the optimized energy was monitored over the last 200 optimization steps as part of the numerical convergence analysis.

These formulas and settings summarize the numerical workflow used in the present manuscript: local Metropolis sampling in the mixed bosonic-$\mathbb{Z}_2$ Hilbert space, variational optimization of the RBM parameters by gradient-based updates or stochastic reconfiguration, and evaluation of the observables from Monte Carlo averages over the optimized neural quantum state.

\section*{Acknowledgements}
G.A.A.V. and R.J.A.R gratefully acknowledge the support of a graduate scholarship provided by the Mexican Ministry of Science, Humanities, Technology, and Innovation (SECIHTI) during the development of this work. I.P.C would like to thank his colleagues at the Instituto de Ciencias Físicas (UNAM, Cuernavaca) for their hospitality and support during a difficult period. In particular, I.P.C. thanks Juan Carlos, Antonio, Luis, and Thomas. This work was partly covered with support from Christof Jung Kohl’s CONAHCYT project (No. 425854).
\end{multicols}

\medline
\begin{multicols}{2}
\bibliographystyle{rmf-style}
\bibliography{biblio}
\end{multicols}

\end{document}